\documentclass[runningheads]{llncs}
\usepackage{graphicx}
\usepackage{amsmath}
\usepackage{isabelle,isabellesym}
\usepackage{pdfsetup}
\usepackage{doi}
\usepackage[strings]{underscore}
\usepackage{enumitem}

\isabellestyle{it}
\begin{document}
\title{A Modular First Formalisation of\\ Combinatorial Design Theory\thanks{The first author is supported by a Cambridge Australia Scholarship and a Cambridge Department of Technology Qualcomm Premium Research Scholarship.
The work is also supported by the ERC Advanced Grant ALEXANDRIA (Project GA 742178)}}
\author{Chelsea Edmonds \and Lawrence Paulson }
%
%
\institute{Department of Computer Science and Technology\\ University of Cambridge, UK\\ \email{cle47@cam.ac.uk} \hspace{0.5em} \email {lp15@cam.ac.uk}}
\maketitle
\begin{abstract}

Combinatorial design theory studies set systems with certain balance and symmetry properties and has applications to computer science and elsewhere. This paper presents a modular approach to formalising designs for the first time using Isabelle and assesses the usability of a locale-centric approach to formalisations of mathematical structures. We demonstrate how locales can be used to specify numerous types of designs and their hierarchy. The resulting library, which is concise and adaptable, includes formal definitions and proofs for many key properties, operations, and theorems on the construction and existence of designs.

\end{abstract}
\section{Introduction}

The formalisation of mathematics is an area of increasing interest, with benefits including verifying correctness, deeper insights into proofs, and automation. This has lead to substantial development of formal mathematical libraries across several different proof assistants covering a notable portion of undergraduate mathematics. However, one area of mathematics that remains underrepresented is combinatorics. In particular, the field of combinatorial design theory has not previously been formalised in any system. 

Combinatorial design theory is the study of systems of finite sets which meet certain balance and symmetry properties. Many results in design theory have been driven by applications to fields such as communications and security, where formal verification is of increasing interest. This paper presents a general formal library for design theory using a modular approach in Isabelle/HOL.

Locales are Isabelle's module system, and are well suited to the problem of managing the complex hierarchy of design classes. While locales have been available in the current form since the early 2000s, they typically have been used sparingly in mathematical contexts, or alongside other tools such as type classes and records. This project presented the opportunity to explore a locale-centric approach to formalising mathematical structures, building on Ballarin's prior work in algebra \cite{ballarinExploringStructureAlgebra2020}, and using ideas from Noschinski's graph theory library\cite{noschinskiGraphLibraryIsabelle2015}. 

We focus on balanced and block designs to define BIBDs, the most extensively studied class of designs, but also explore how easy it is to extend the formalisation to other design classes and graph theory. Our library includes the formal definitions for many key properties and operations on designs generally. It also explores the formal proof process for theorems on the construction and existence of designs with certain parameters, two basic questions in design theory. 

This paper begins with (2) the necessary background on design theory and locales, then presents (3) the formalisation of fundamental concepts on designs, followed by (4) the development of the BIBD locale hierarchy and (5) extending the formalisation beyond BIBDs. We conclude (6) with a discussion on the locale-centric approach to formalising mathematical structures.

\section{Background}

\subsection{Mathematical Background}

Designs are one of many different combinatorial structures which have emerged in the last century. Formally, a design is defined as follows \cite{stinsonCombinatorialDesignsConstructions2004}:
\begin{definition}[Design]
    A \emph{design} is a pair $(V, B)$ where $V$ is a (finite) set of \emph{points} and $B$ is a (finite) collection of non-empty subsets of $V$ called \emph{blocks}.
\end{definition}

Designs are also referred to as \textit{incidence structures} \cite{bethDesignTheory1999a} and more specifically, \textit{incidence set systems}. There are four sets defined on key set system properties which can be restricted to impose structural conditions on a design \cite{colbournHandbookCombinatorialDesigns2007}. 
\begin{enumerate}[label=\roman*)]
    \item The set $K$ of all block sizes in the design.
    \item The set $R$ of replication numbers for points in the design, where the \textit{point replication number} $r_x$ is the number of blocks the point $x$ occurs in.
    \item The set $\Lambda_t$ of $t$-indices for $t \ge 0$ the design. For any $t$ subset of points, the $t$ \textit{points index} is the number of blocks that subset occurs in.
    \item The set $I$ of intersection numbers. For two blocks in a design, the \textit{intersection number} is the number of points the blocks intersect on.
\end{enumerate}

Using different structural restrictions results in numerous classes of designs. The designs of most interest mathematically usually involve the combination of several restrictions, such as \textit{balanced incomplete block designs (BIBD)}. 
\begin{definition}[BIBD]
    Let $v, k$, and $\lambda$, be positive integers such that $2 \le k < v$. A \emph{$(v, k, \lambda)$-design} is a design with $v$ points where every block has $k$ elements and where every pair of points occurs in exactly $\lambda$ blocks.
\end{definition}

The balance and uniformity properties of a BIBD, as well as properties like resolvability and symmetry, lead to further design variations such as group divisible designs (GDDs), pairwise balanced designs (PBDs), triple systems, and resolvable designs \cite{colbournHandbookCombinatorialDesigns2007}. 

Most open questions in design theory concern either the existence of a design with certain parameters or the construction of certain designs for which existence is already known \cite{stinsonCombinatorialDesignsConstructions2004}. Numerous operations have been defined to reason on the construction of designs, several of which this paper explores. Proofs in design theory often draw on other fields of mathematics, and combinatorial counting techniques, which present interesting formalisation challenges.

Set systems are the underlying construct of a design, and are the basis for numerous other structures such as hypergraphs, matrices, geometries, codes, and graphs \cite{colbournHandbookCombinatorialDesigns2007}. As such, designs have close links to these fields, and they are often used in proofs on designs. For example, it can be seen that an undirected simple graph is a design, where the vertices are points and edges are 2-blocks. The design of a graph is normally not interesting from a design theoretic standpoint, as it often lacks the structure of many design classes. However, a $r$-regular graph can be thought of as a design with replication number $r$. Graphs are also useful for representing other design properties such as resolvability \cite{cameronDesignsGraphsCodes1996}.

\subsection{Isabelle and Locales}

Isabelle/HOL is an interactive proof assistant built on higher order logic \cite{paulsonComputationalLogicIts2018}. It has extensive libraries of formalised mathematics, including the largest number of results related to combinatorics from a survey of several proof assistants. These libraries, combined with powerful built-in tools such as the Isar proof language and Sledgehammar, make Isabelle an ideal choice for this formalisation work. 

Locales are an important extension of the Isar proof language. They act as a module system within Isabelle, providing persistent contexts which can be used across numerous theories drawing on similar structures \cite{ballarinLocalesLocaleExpressions2004}. In the simplest form, a locale declaration introduces parameters and assumptions. Each parameter has a specified type and can even have associated syntax. Once defined, a locale can be extended with definitions, notation and theorems within its context.

Locale expressions were designed to support multiple inheritance and thus offer extensive flexibility. Existing locales can be combined to create a new locale and extended by adding new parameters and assumptions \cite{ballarinLocalesLocaleExpressions2004}. The locale hierarchy can be transformed using the \textbf{sublocale} command, which is used to show indirect inheritance between two separately specified locales. It is also possible to instantiate locale parameters and instances through locale expressions and interpretations. A full tutorial introduction on locales is available with Isabelle~\cite{ballarinTutorialLocalesLocale2010a}.

\section{The Basic Design Formalisation}

Formalising design theory presents a number of initial challenges. Of particular note is (i) notation and definition inconsistencies in the literature, (ii) the significant number of definitions and properties, and (iii) the complex relations between different classes of designs, as well as other combinatorial structures. 

To narrow the focus of the formalisation, addressing (ii), initial formalisation efforts focussed on defining BIBDs and operations commonly found in computational libraries for designs such as GAP \cite{soicherDesignsGroupsComputing2013}. Proofs focused on enabling reasoning on common design properties, constructions, and existence requirements.

To address (i), key decisions were made early in the formalisation process covered below and in Sect.\ts4. For consistency, the Handbook of Combinatorial Designs was the primary reference for definitions, with publications from well known researchers such as Stinson \cite{stinsonCombinatorialDesignsConstructions2004} serving as alternatives  when needed. Challenge (iii) is the motivation for our locale-centric approach to formalising fundamental definitions and operations for general designs, discussed below.

\subsection{Pre-designs}

First, a locale representing a general incidence system is defined, which introduces the core components of a design: a block collection formalised using multisets, a point set, and a well-formed assumption: 

\medskip\isamarkuptrue%
\isacommand{locale}\isamarkupfalse%
\ incidence{\isacharunderscore}{\kern0pt}system\ {\isacharequal}{\kern0pt}\ \isanewline
\ \ \isakeyword{fixes}\ point{\isacharunderscore}{\kern0pt}set\ {\isacharcolon}{\kern0pt}{\isacharcolon}{\kern0pt}\ {\isachardoublequoteopen}{\isacharprime}{\kern0pt}a\ set{\isachardoublequoteclose}\ {\isacharparenleft}{\kern0pt}{\isachardoublequoteopen}{\isasymV}{\isachardoublequoteclose}{\isacharparenright}{\kern0pt} \isakeyword{and}\ block{\isacharunderscore}{\kern0pt}collection\ {\isacharcolon}{\kern0pt}{\isacharcolon}{\kern0pt}\ {\isachardoublequoteopen}{\isacharprime}{\kern0pt}a\ set\ multiset{\isachardoublequoteclose}\ {\isacharparenleft}{\kern0pt}{\isachardoublequoteopen}{\isasymB}{\isachardoublequoteclose}{\isacharparenright}{\kern0pt}\isanewline
\ \ \isakeyword{assumes}\ wellformed{\isacharcolon}{\kern0pt}\ {\isachardoublequoteopen}b\ {\isasymin}{\isacharhash}{\kern0pt}\ {\isasymB}\ {\isasymLongrightarrow}\ b\ {\isasymsubseteq}\ {\isasymV}{\isachardoublequoteclose}\medskip

Definition 1 (see Sect. \ts 2.1) states designs are finite, which is added as an assumption in the \textit{finite-incidence-system} locale. Lastly, a design often has the additional condition that blocks must be non-empty \cite{stinsonCombinatorialDesignsConstructions2004}:

\medskip\isamarkuptrue%
\isacommand{locale}\isamarkupfalse%
\ design\ {\isacharequal}{\kern0pt}\ finite{\isacharunderscore}{\kern0pt}incidence{\isacharunderscore}{\kern0pt}system\ {\isacharplus}{\kern0pt}\isanewline
\ \ \isakeyword{assumes}\ blocks{\isacharunderscore}{\kern0pt}nempty{\isacharcolon}{\kern0pt}\ {\isachardoublequoteopen}bl\ {\isasymin}{\isacharhash}{\kern0pt}\ {\isasymB}\ {\isasymLongrightarrow}\ bl\ {\isasymnoteq}\ {\isacharbraceleft}{\kern0pt}{\isacharbraceright}{\kern0pt}{\isachardoublequoteclose}\medskip

Some design definitions further impose the condition that a design must be non-empty \cite{soicherDesignsGroupsComputing2013}. This is important for some classes of designs, but constrains others unnecessarily, and hence is defined separately in the locale \textit{proper-designs}. 

\subsection{Basic Design Properties}
The four key properties on elements of a set system are block size, intersection numbers, point indices, and replication numbers. These are defined outside of a locale context, as they are properties on components of the set system, rather than the entire structure. The definition of the points index property is below:

\medskip\isamarkuptrue%
\isacommand{definition}\isamarkupfalse%
\ points{\isacharunderscore}{\kern0pt}index\ {\isacharcolon}{\kern0pt}{\isacharcolon}{\kern0pt}\ {\isachardoublequoteopen}{\isacharprime}{\kern0pt}a\ set\ multiset\ {\isasymRightarrow}\ {\isacharprime}{\kern0pt}a\ set\ {\isasymRightarrow}\ nat{\isachardoublequoteclose}\ \isakeyword{where}\isanewline
{\isachardoublequoteopen}points{\isacharunderscore}{\kern0pt}index\ B\ ps\ {\isasymequiv}\ size\ {\isacharbraceleft}{\kern0pt}{\isacharhash}{\kern0pt}b\ {\isasymin}{\isacharhash}{\kern0pt}\ B\ {\isachardot}{\kern0pt}\ ps\ {\isasymsubseteq}\ b{\isacharhash}{\kern0pt}{\isacharbraceright}{\kern0pt}{\isachardoublequoteclose}\medskip

Numerous lemmas for reasoning on these properties can be defined in the context of incidence systems and designs. Using these properties, the four key sets outlined in (2) can be defined within the general \textit{incidence_system} locale.  The definition of the point indices set is given below:

\medskip\isacommand{definition}\isamarkupfalse%
\ point{\isacharunderscore}{\kern0pt}indices\ {\isacharcolon}{\kern0pt}{\isacharcolon}{\kern0pt}\ {\isachardoublequoteopen}int\ {\isasymRightarrow}\ int\ set{\isachardoublequoteclose}\ \isakeyword{where}\isanewline
{\isachardoublequoteopen}point{\isacharunderscore}{\kern0pt}indices\ t\ {\isasymequiv}\ {\isacharbraceleft}{\kern0pt}\ points{\isacharunderscore}{\kern0pt}index\ {\isasymB}\ ps\ {\isacharbar}{\kern0pt}\ ps{\isachardot}{\kern0pt}\ int\ {\isacharparenleft}{\kern0pt}card\ ps{\isacharparenright}{\kern0pt}\ {\isacharequal}{\kern0pt}\ t\ {\isasymand}\ ps\ {\isasymsubseteq}\ {\isasymV}{\isacharbraceright}{\kern0pt}{\isachardoublequoteclose}\medskip

Lastly, the basic design locale includes a number of abbreviations to mirror terminology in the literature: design supports, multiplicity of blocks, incomplete blocks, design order $v$ (number of points), and design size $b$ (number of blocks). The multiplicity and design support abbreviations are used to establish a new locale for \textit{simple-designs}, where block multiplicity is at most 1.

\subsection{Basic Design Operations}

Designs are often constructed by building on pre-existing designs through operations. The three main operations considered for the formalisation are design complements, multiples, and combinations. The \textit{complement} of a design $(V, B)$ is the design $(V, \{V - bl . bl \in B\})$, where $V - bl$ is the \textit{block complement} of the block $bl$. A \textit{multiple} of a design multiplies the block multiset by some constant $n \ge 0$, and \textit{combining} designs is simply the union of the point set and addition of the block multisets. The formal definitions for these operations are defined within the incidence system locale, such as the complement operation below, along with a number of relevant lemmas.

\medskip\isacommand{definition}\isamarkupfalse%
\ complement{\isacharunderscore}{\kern0pt}blocks\ {\isacharcolon}{\kern0pt}{\isacharcolon}{\kern0pt}\ {\isachardoublequoteopen}{\isacharprime}{\kern0pt}a\ set\ multiset{\isachardoublequoteclose}\ \isakeyword{where}\isanewline
{\isachardoublequoteopen}complement{\isacharunderscore}{\kern0pt}blocks\ {\isasymequiv}\ {\isacharbraceleft}{\kern0pt}{\isacharhash}{\kern0pt}\ block{\isacharunderscore}{\kern0pt}complement\ bl\ {\isachardot}{\kern0pt}\ bl\ {\isasymin}{\isacharhash}{\kern0pt}\ {\isasymB}\ {\isacharhash}{\kern0pt}{\isacharbraceright}{\kern0pt}{\isachardoublequoteclose}\medskip

Numerous basic lemmas are shown for all three operations. In particular, \textit{multiple} and \textit{combine} are shown to be closed under the design conditions, and \textit{complement} will result in a design if the original blocks are incomplete. We additionally formalised a number of simple computational operations, such as addition and deletion of points, which are useful when constructing new designs. 

\section{The Block Design Hierarchy}

By definition 2, a BIBD could be easily defined in a single locale with parameters for block size, index, and replication number, as well as assumptions on balance, constant replication, and uniformity conditions. However, this approach would have significant limitations. Although a replication number is widely used in proofs of a BIBD, its value is implied by the other parameters, hence the assumption is unnecessary. Additionally, this could result in a significant amount of rework if more general designs than BIBDs need to be formalised.

The approach taken in this formalisation uses the idea of little and tiny theories \cite{farmerLittleTheories1992}\cite{caretteMathSchemeLibraryPreliminary2011}, discussed in Sect.\ts6. Each locale definition adds a single concept, and lemmas on properties and operations are introduced in the most general locale possible. This section explores the process of building up the locale hierarchy to BIBDs through the gradual specification of more general locales. 

\subsection{Restricting Block Size}

The first new parameter in a BIBD is $k$, the uniform size of a design's blocks. Formally, it is introduced through the block design locale:

\medskip\isacommand{locale}\isamarkupfalse%
\ block{\isacharunderscore}{\kern0pt}design\ {\isacharequal}{\kern0pt}\ proper{\isacharunderscore}{\kern0pt}design\ {\isacharplus}{\kern0pt}\ \isanewline
\ \ \isakeyword{fixes}\ u{\isacharunderscore}{\kern0pt}block{\isacharunderscore}{\kern0pt}size\ {\isacharcolon}{\kern0pt}{\isacharcolon}{\kern0pt}\ int\ {\isacharparenleft}{\kern0pt}{\isachardoublequoteopen}{\isasymk}{\isachardoublequoteclose}{\isacharparenright}{\kern0pt}\isanewline
\ \ \isakeyword{assumes}\ uniform\ {\isacharbrackleft}{\kern0pt}simp{\isacharbrackright}{\kern0pt}{\isacharcolon}{\kern0pt}\ {\isachardoublequoteopen}bl\ {\isasymin}{\isacharhash}{\kern0pt}\ {\isasymB}\ {\isasymLongrightarrow}\ block{\isacharunderscore}{\kern0pt}size\ bl\ {\isacharequal}{\kern0pt}\ {\isasymk}{\isachardoublequoteclose}\medskip

A key design decision was to let uniform parameters such as block size be integers. While these are clearly positive and could be natural numbers, proofs often require manipulating algebraic expressions involving subtraction on the parameters, which is notably simpler to do using integers in Isabelle.

A number of lemmas are defined within the \textit{block_design} locale. Recurring themes on proofs throughout the formalisation include proving inequality relationships on parameters, such as $k \le v$, and that the three main operations defined in (3.3) result in another type of this design given certain conditions. For a block design, \textit{multiple} and \textit{combine} are clearly closed, whereas \textit{complement} requires an additional assumption. Two main proof strategies are used for these lemmas: a direct proof using introduction rules, and the more expressive \textbf{interpret} proof structure, discussed in Sect.\ts6.4.

A \textit{$K$-design} is a generalisation of a $k$-design which limits the size of blocks to a finite set of positive integers. An important specialisation of a block design is an \textit{incomplete design} where all blocks are incomplete, i.e. $k < v$. 

\subsection{Balanced Designs}

The balance property and its variations are widely used across different design classes. The most general balanced design is a $t$-wise balanced design or tBD, where for some $1 \le t \le v$, the points index of a $t$-sized subset of points equals~$\lambda_t$.

\medskip\isacommand{locale}\isamarkupfalse%
\ twise{\isacharunderscore}{\kern0pt}balance\ {\isacharequal}{\kern0pt}\ proper{\isacharunderscore}{\kern0pt}design\ {\isacharplus}{\kern0pt}\ \isanewline
\ \ \isakeyword{fixes}\ grouping\ {\isacharcolon}{\kern0pt}{\isacharcolon}{\kern0pt}\ int\ {\isacharparenleft}{\kern0pt}{\isachardoublequoteopen}{\isasymt}{\isachardoublequoteclose}{\isacharparenright}{\kern0pt}\ \isakeyword{and}\ index\ {\isacharcolon}{\kern0pt}{\isacharcolon}{\kern0pt}\ int\ {\isacharparenleft}{\kern0pt}{\isachardoublequoteopen}{\isasymLambda}\isactrlsub t{\isachardoublequoteclose}{\isacharparenright}{\kern0pt}\isanewline
\ \ \isakeyword{assumes}\ t{\isacharunderscore}{\kern0pt}non{\isacharunderscore}{\kern0pt}zero{\isacharcolon}{\kern0pt}\ {\isachardoublequoteopen}{\isasymt}\ {\isasymge}\ {\isadigit{1}}{\isachardoublequoteclose}\ \isakeyword{and}\ t{\isacharunderscore}{\kern0pt}lt{\isacharunderscore}{\kern0pt}order{\isacharcolon}{\kern0pt}\ {\isachardoublequoteopen}{\isasymt}\ {\isasymle}\ {\isasymv}{\isachardoublequoteclose}\isanewline
\ \ \isakeyword{and}\ balanced\ {\isacharbrackleft}{\kern0pt}simp{\isacharbrackright}{\kern0pt}{\isacharcolon}{\kern0pt}\ {\isachardoublequoteopen}ps\ {\isasymsubseteq}\ {\isasymV}\ {\isasymLongrightarrow}\ card\ ps\ {\isacharequal}{\kern0pt}\ {\isasymt}\ {\isasymLongrightarrow}\ points{\isacharunderscore}{\kern0pt}index\ {\isasymB}\ ps\ {\isacharequal}{\kern0pt}\ {\isasymLambda}\isactrlsub t{\isachardoublequoteclose}\medskip

Note that as $\lambda$ is reserved in Isabelle, $\Lambda$ is used in its place. Also, as the parameters $t$ and $\lambda_t$ and their assumptions are linked, there is no sensible way to further break down the locale. Within the locale context it can easily be shown that combining two designs with the same point set, or applying the multiple operation, results in another tBD\@. A $t$-wise balanced design can include a set $K$ of valid block sizes, which is formalised by combining the tBD and $K$ block design locales.

BIBDs are interested in pairwise balance, where $t = 2$. A PBD is a clear specialisation of a tBD which can be defined formally using the \textit{for} command in a locale definition to instantiate one parameter and simplify syntax.

\medskip\isacommand{locale}\isamarkupfalse%
\ pairwise{\isacharunderscore}{\kern0pt}balance\ {\isacharequal}{\kern0pt}\ t{\isacharunderscore}wise{\isacharunderscore}{\kern0pt}balance\ {\isasymV}\ {\isasymB}\ {\isadigit{2}}\ {\isasymLambda}\ \isanewline \ \isakeyword{for}\ point{\isacharunderscore}{\kern0pt}set\ {\isacharparenleft}{\kern0pt}{\isachardoublequoteopen}{\isasymV}{\isachardoublequoteclose}{\isacharparenright}{\kern0pt}\ \isakeyword{and}\ block{\isacharunderscore}{\kern0pt}collection\ {\isacharparenleft}{\kern0pt}{\isachardoublequoteopen}{\isasymB}{\isachardoublequoteclose}{\isacharparenright}{\kern0pt}\ \isakeyword{and}\ index\ {\isacharparenleft}{\kern0pt}{\isachardoublequoteopen}{\isasymLambda}{\isachardoublequoteclose}{\isacharparenright}{\kern0pt}\medskip

There are several variations on PBDs in the literature depending on block size properties and the value of $\lambda$, which are easy to specify by combining locales and the use of the sublocale declaration, following the functor proof pattern \cite{ballarinExploringStructureAlgebra2020}. 

\subsection{t-designs}
An important generalisation of BIBDs are $t$-designs. Given the modular structure of the existing locale declarations, they can be easily specified by combining locales on incomplete block designs and $t$-wise balanced designs. Additionally, an extra assumption is required on the relationship of the parameters $t$ and $k$.

\medskip\isacommand{locale}\isamarkupfalse%
\ tdesign\ {\isacharequal}{\kern0pt}\ incomplete{\isacharunderscore}{\kern0pt}design\ {\isacharplus}{\kern0pt}\ t{\isacharunderscore}wise{\isacharunderscore}{\kern0pt}balance\ {\isacharplus}{\kern0pt}\ \isanewline
\ \ \isakeyword{assumes}\ block{\isacharunderscore}{\kern0pt}size{\isacharunderscore}{\kern0pt}t{\isacharcolon}{\kern0pt}\ {\isachardoublequoteopen}{\isasymt}\ {\isasymle}\ {\isasymk}{\isachardoublequoteclose}\medskip

In addition to $t$-designs, the related concepts of $t$-covering and $t$-packing designs are also formalised, where $\lambda_t$ has a slightly different meaning, a typical example of design notation inconsistencies. A $t$-covering design is a relaxed version of a tBD where, for all point subsets of size $t$, $\lambda_t$ is a lower bound on the points index. A $t$-packing design mirrors this with an upper bound. Given the different meaning of the parameter $\lambda_t$, these designs build only on block designs. If a design is incomplete, $t$-packing and $t$-covering, then it is a $t$-design.

Additionally, a locale is declared for \textit{Steiner systems}: $t$-designs where $\lambda_t = 1$. Then it can be proven that all blocks in a Steiner system have a multiplicity of 1. Hence it can be shown that Steiner systems are simple designs using sublocales.

\subsection{Uniform Replication Number}
When every point in a design has the same replication number, $r$ is known as the replication number of the design.

\medskip\isacommand{locale}\isamarkupfalse%
\ constant{\isacharunderscore}{\kern0pt}rep{\isacharunderscore}{\kern0pt}design\ {\isacharequal}{\kern0pt}\ proper{\isacharunderscore}{\kern0pt}design\ {\isacharplus}{\kern0pt}\isanewline
\ \ \isakeyword{fixes}\ design{\isacharunderscore}{\kern0pt}rep{\isacharunderscore}{\kern0pt}number\ {\isacharcolon}{\kern0pt}{\isacharcolon}{\kern0pt}\ int\ {\isacharparenleft}{\kern0pt}{\isachardoublequoteopen}{\isasymr}{\isachardoublequoteclose}{\isacharparenright}{\kern0pt}\isanewline
\ \ \isakeyword{assumes}\ rep{\isacharunderscore}{\kern0pt}number\ {\isacharbrackleft}{\kern0pt}simp{\isacharbrackright}{\kern0pt}{\isacharcolon}{\kern0pt}\ {\isachardoublequoteopen}x\ {\isasymin}\ {\isasymV}\ {\isasymLongrightarrow}\ \ {\isasymB}\ rep\ x\ {\isacharequal}{\kern0pt}\ {\isasymr}{\isachardoublequoteclose}\medskip

As with the other locales, we can prove that $r > 0$, and that the complement, multiple, and combination operators result in another constant replication design under certain conditions within the locale's context.

\subsection{BIBDs and Proofs}
The final BIBD locale declaration builds on the $t$-design locale and is now simple to define using the \textbf{for} command to instantiate $t = 2$, as with PBDs. 

\medskip\isacommand{locale}\isamarkupfalse%
\ bibd\ {\isacharequal}{\kern0pt}\ t{\isacharunderscore}design\ {\isasymV}\ {\isasymB}\ {\isasymk}\ {\isadigit{2}}\ {\isasymLambda}\ \isakeyword{for}\ point{\isacharunderscore}{\kern0pt}set\ {\isacharparenleft}{\kern0pt}{\isachardoublequoteopen}{\isasymV}{\isachardoublequoteclose}{\isacharparenright}{\kern0pt}\ \isakeyword{and}\ block{\isacharunderscore}{\kern0pt}collection\ {\isacharparenleft}{\kern0pt}{\isachardoublequoteopen}{\isasymB}{\isachardoublequoteclose}{\isacharparenright}{\kern0pt}\isanewline \ \ \ \ \isakeyword{and}\ u{\isacharunderscore}{\kern0pt}block{\isacharunderscore}{\kern0pt}size\ {\isacharparenleft}{\kern0pt}{\isachardoublequoteopen}{\isasymk}{\isachardoublequoteclose}{\isacharparenright}{\kern0pt}\ \isakeyword{and}\ index\ {\isacharparenleft}{\kern0pt}{\isachardoublequoteopen}{\isasymLambda}{\isachardoublequoteclose}{\isacharparenright}\medskip

Figure 1 gives an overview of the final locale hierarchy for BIBDs, with sublocale relationships represented by a dotted line. Using this structure, we used BIBDs as case study for doing more involved proofs on both existence and construction. Many of these proofs required formalising a counting proof, the full details of which are out of scope of this paper.
\begin{figure}[t]
    \includegraphics[width = \textwidth]{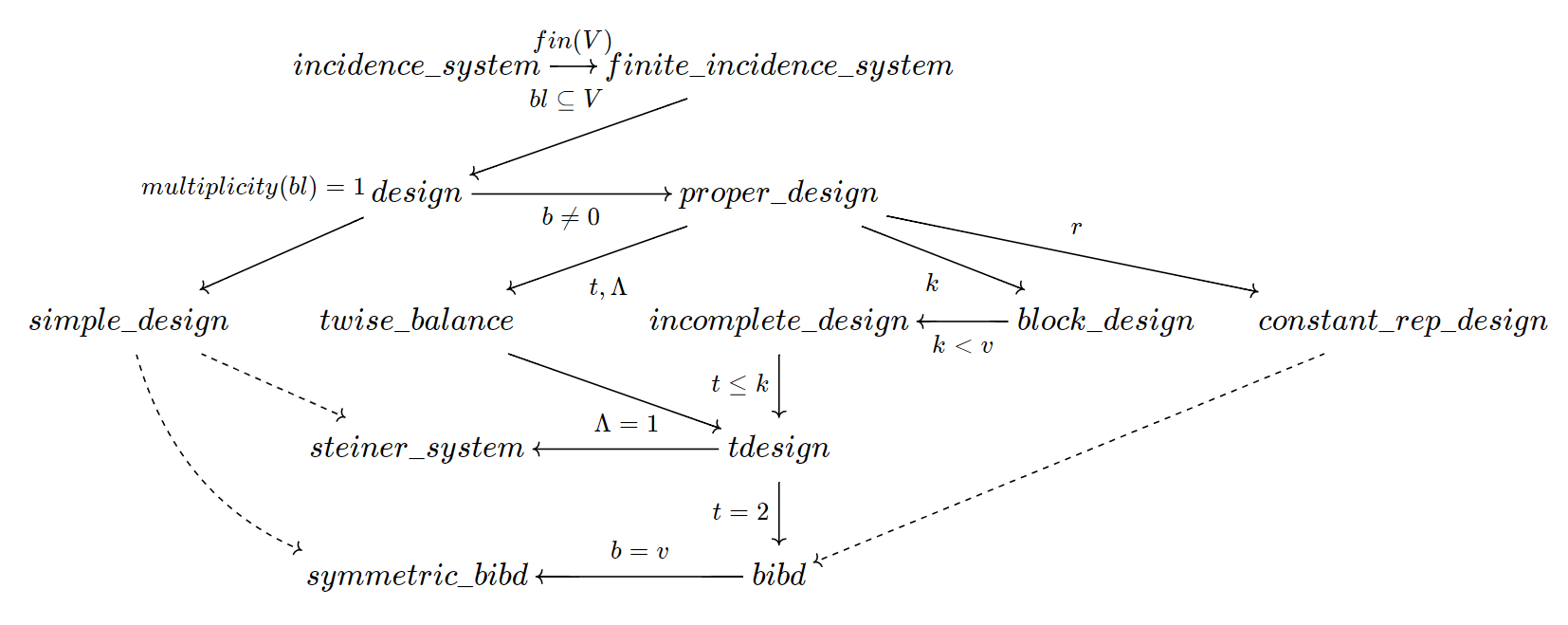}
    \caption{The BIBD Locale Hierarchy} \label{fig1}
\end{figure}
    
There are two necessary conditions on BIBD existence, which therefore must hold in the locale context. These define important relationships between parameters:   $r  (k - 1) = \lambda (v - 1)$ and  $v r = b k$. Notably, this uses the design replication number, which is not yet defined in the BIBD context. However, the first condition can still be shown to hold for each point's replication number $r_x$, which in turn proves $r$ is constant. This results in the following sublocale declaration.

\medskip
\isamarkuptrue%
\isacommand{sublocale}\isamarkupfalse%
\ bibd\ {\isasymsubseteq}\ constant{\isacharunderscore}{\kern0pt}rep{\isacharunderscore}{\kern0pt}design\ {\isasymV}\ {\isasymB}\ \ {\isachardoublequoteopen}{\isacharparenleft}{\kern0pt}{\isasymLambda}\ {\isacharasterisk}{\kern0pt}\ {\isacharparenleft}{\kern0pt}{\isasymv}\ {\isacharminus}{\kern0pt}\ {\isadigit{1}}{\isacharparenright}{\kern0pt}\ div\ {\isacharparenleft}{\kern0pt}{\isasymk}\ {\isacharminus}{\kern0pt}\ {\isadigit{1}}{\isacharparenright}{\kern0pt}{\isacharparenright}{\kern0pt}{\isachardoublequoteclose}\isanewline
\ \ \isacommand{using}\isamarkupfalse%
\ r{\isacharunderscore}{\kern0pt}constant{\isacharunderscore}{\kern0pt}{\isadigit{2}}\ \isacommand{by}\isamarkupfalse%
\ {\isacharparenleft}{\kern0pt}unfold{\isacharunderscore}{\kern0pt}locales{\isacharparenright}{\kern0pt}\ simp{\isacharunderscore}{\kern0pt}all%

\medskip

These necessary conditions enable proofs of useful lemmas on inequalities between parameters, and set up the formalisation for further construction proofs.

As with previous locales, it is simple to prove the combination and multiple operations result in another BIBD with simply defined parameters assuming equal point sets. The complement of a $(v, k, \lambda)$-design is a $(v, v - k, b + \lambda - 2r)$-design. These parameters are more complicated and so are their proofs. The final proof for the main \textit{complement-bibd} lemma is a good example of how constructive design proofs can be presented with little effort using interpretation and the Isar proof language (see Sect.\ts6).

\subsection{BIBD Extensions}

Symmetric BIBDs are an extension of BIBDs where $b = v$, as shown in Fig.\ts\ref{fig1}. An important theorem on symmetric designs is the \textit{intersection property}: the intersection number of any two blocks in the design is equal to the design index $\lambda$. We have formalised its delicate counting proof, making use of the necessary conditions on a BIBD\@. 

The BIBD locale also includes definitions and lemmas on residual and derived designs, which are common constructions specific to BIBDs. The formal definitions of these operations resolve some ambiguities in the literature which use set comprehensions and notation to describe operations on multisets. Using the intersection property, it is possible to prove that the derived and residual designs of a symmetric BIBD are also BIBDs. The intersection property and sublocale command can also be used to show that symmetric designs are simple. 

\section{Extending the Formalisation}

This section investigates the ease of extending the formalisation to a number of other structures in design theory and graph theory.

\subsection{Resolvable Designs}

A \textit{resolution class} of a design is a partition of the point set using blocks. A partition of the blocks into resolution classes is known as a \textit{resolution}, and a design with a resolution is \textit{resolvable}. 
While set partitions are well covered in Isabelle, we had to formalise multiset partitions. The concepts of a resolution class and resolution were then easily defined within \textit{incidence-system}. A resolvable design is represented by a new locale building on designs: 

\medskip\isacommand{locale}\isamarkupfalse%
\ resolvable{\isacharunderscore}{\kern0pt}design\ {\isacharequal}{\kern0pt}\ design\ {\isacharplus}{\kern0pt}\ \isanewline
\ \ \isakeyword{fixes}\ partition\ {\isacharcolon}{\kern0pt}{\isacharcolon}{\kern0pt}\ {\isachardoublequoteopen}{\isacharprime}{\kern0pt}a\ set\ multiset\ multiset{\isachardoublequoteclose}\ {\isacharparenleft}{\kern0pt}{\isachardoublequoteopen}{\isasymP}{\isachardoublequoteclose}{\isacharparenright}{\kern0pt}\isanewline
\ \ \isakeyword{assumes}\ resolvable{\isacharcolon}{\kern0pt}\ {\isachardoublequoteopen}resolution\ {\isasymP}{\isachardoublequoteclose}\medskip

Further classes of resolvable designs were defined by combining this locale with block designs and BIBDs. The resolvable specification enables us to prove a number of new relations between the parameters of these designs, such as $k | v$ in a resolvable block design. A proof was also completed for an alternate statement of Bose's inequality on resolvable BIBDs based on Stinson's approach~\cite{stinsonCombinatorialDesignsConstructions2004}.

\subsection{Group Divisible Designs}

GDDs are closely related to PBDs and are often studied simultaneously. As such, they were an ideal case study for extending the BIBD hierarchy. A GDD is a design which has a non-empty group $G$ which partitions the point set, and a points index of $\lambda$ or 0 for each pair depending on if points occurs together in $G$.

Continuing with the little theories approach, the definition is split into two locales. Firstly, a \textit{group-design} locale is declared, which introduces the parameter $G$ and the partition assumption. Within this locale a number of properties of the group in GDDs are defined. This includes the concept of group types, which represent a GDDs structure by the size of the sets in $G$. A GDD locale then introduces the index parameter and assumptions: 

\medskip\isacommand{locale}\isamarkupfalse%
\ GDD\ {\isacharequal}{\kern0pt}\ group{\isacharunderscore}{\kern0pt}design\ {\isacharplus}{\kern0pt}\ \isakeyword{fixes}\ index\ {\isacharcolon}{\kern0pt}{\isacharcolon}{\kern0pt}\ int\ {\isacharparenleft}{\kern0pt}{\isachardoublequoteopen}{\isasymLambda}{\isachardoublequoteclose}{\isacharparenright}{\kern0pt}\isanewline
\ \ \isakeyword{assumes}\ index{\isacharunderscore}{\kern0pt}ge{\isacharunderscore}{\kern0pt}{\isadigit{1}}{\isacharcolon}{\kern0pt}\ {\isachardoublequoteopen}{\isasymLambda}\ {\isasymge}\ {\isadigit{1}}{\isachardoublequoteclose}\isanewline
\ \ \isakeyword{assumes}\ index{\isacharunderscore}{\kern0pt}together{\isacharcolon}{\kern0pt}\ {\isachardoublequoteopen}\textlbrackdbl G\ {\isasymin}\ {\isasymG};\ x\ {\isasymin}\ G;\ y\ {\isasymin}\ G; \ x\ {\isasymnoteq}\ y\textrbrackdbl\ {\isasymLongrightarrow}\ points{\isacharunderscore}{\kern0pt}index\isanewline \ \ \ \ {\isasymB}\ {\isacharbraceleft}{\kern0pt}x{\isacharcomma}{\kern0pt}\ y{\isacharbraceright}{\kern0pt}\ {\isacharequal}{\kern0pt}\ {\isadigit{0}}{\isachardoublequoteclose}\ \isakeyword{and}\ index{\isacharunderscore}{\kern0pt}distinct{\isacharcolon}{\kern0pt}\ {\isachardoublequoteopen}\textlbrackdbl G{\isadigit{1}}\ {\isasymin}\ {\isasymG};\ G{\isadigit{2}}\ {\isasymin}\ {\isasymG};\ G{\isadigit{1}}\ {\isasymnoteq}\ G{\isadigit{2}};\ x\ {\isasymin}\ G{\isadigit{1}};\isanewline \ \ \ \ y\ {\isasymin}\ G{\isadigit{2}}\textrbrackdbl\ {\isasymLongrightarrow}\ points{\isacharunderscore}{\kern0pt}index\ {\isasymB}\ {\isacharbraceleft}{\kern0pt}x{\isacharcomma}{\kern0pt}\ y{\isacharbraceright}{\kern0pt}\ {\isacharequal}{\kern0pt}\ {\isasymLambda}{\isachardoublequoteclose}\medskip

As with PBDs, GDDs are defined in different ways, commonly combined with $K$ block designs, or certain instantiated parameters, which can easily be formalised using locales. Operations such as adding and deleting points, or combining the group sets and blocks are common on both PBDs and GDDs. For example, combining the group of a $K$-GDD with its blocks results in a PBD with the same point set, a block collection containing both groups and blocks of the original GDD, and a size set $K$. Authors often use these constructions without proofs and lacking necessary assumptions.

\subsection{Design Isomorphisms}

Two designs $(V, B)$ and $(V', B')$ are \textit{isomorphic} if there exists a bijection $\pi$ such that $V' = \pi(V)$ and $B' = \{\pi (bl) . bl \in B'\}$. There are two obvious ways of formalising this relation: through a number of definitions, or through another locale. The second approach enables direct and concise reasoning on an isomorphism relation by using two labelled instances of the same locale:

\medskip\isacommand{locale}\isamarkupfalse%
\ incidence{\isacharunderscore}{\kern0pt}system{\isacharunderscore}{\kern0pt}isomorphism\ {\isacharequal}{\kern0pt}\ source{\isacharcolon}{\kern0pt}\ incidence{\isacharunderscore}{\kern0pt}system\ {\isasymV}\ {\isasymB}\ {\isacharplus}{\kern0pt}\ target{\isacharcolon}{\kern0pt}\isanewline incidence{\isacharunderscore}{\kern0pt}system\ {\isasymV}{\isacharprime}{\kern0pt}\ {\isasymB}{\isacharprime}{\kern0pt}\ \isakeyword{for}\ {\isachardoublequoteopen}{\isasymV}{\isachardoublequoteclose}\ \isakeyword{and}\ {\isachardoublequoteopen}{\isasymB}{\isachardoublequoteclose}\ \isakeyword{and}\ {\isachardoublequoteopen}{\isasymV}{\isacharprime}{\kern0pt}{\isachardoublequoteclose}\ \isakeyword{and}\ {\isachardoublequoteopen}{\isasymB}{\isacharprime}{\kern0pt}{\isachardoublequoteclose}\ {\isacharplus}{\kern0pt}\ \isakeyword{fixes}\ bij{\isacharunderscore}{\kern0pt}map\ {\isacharparenleft}{\kern0pt}{\isachardoublequoteopen}{\isasympi}{\isachardoublequoteclose}{\isacharparenright}{\kern0pt}\isanewline
\ \ \isakeyword{assumes}\ bij{\isacharcolon}{\kern0pt}\ {\isachardoublequoteopen}bij{\isacharunderscore}{\kern0pt}betw\ {\isasympi}\ {\isasymV}\ {\isasymV}{\isacharprime}{\kern0pt}{\isachardoublequoteclose}\ \isakeyword{and}\ block{\isacharunderscore}{\kern0pt}img{\isacharcolon}{\kern0pt}\ {\isachardoublequoteopen}image{\isacharunderscore}{\kern0pt}mset\ {\isacharparenleft}{\kern0pt}{\isacharparenleft}{\kern0pt}{\isacharbackquote}{\kern0pt}{\isacharparenright}{\kern0pt}\ {\isasympi}{\isacharparenright}{\kern0pt}\ {\isasymB}\ {\isacharequal}{\kern0pt}\ {\isasymB}{\isacharprime}{\kern0pt}{\isachardoublequoteclose}\medskip

Within the locale, it is easy to show how elements in $(V, B)$ map to $(V', B')$, and that $\pi^{-1}$ also defines an isomorphic relation. Furthermore, by extending the locale to design instances, the four key properties on set systems are proven to be identical for isomorphic designs. Even with a locale approach, it is still easy to work with isomorphisms outside of the locale if required: below, we define the concept of isomorphic designs on set systems using the locale definition.

\medskip\isacommand{definition}\isamarkupfalse%
\ isomorphic{\isacharunderscore}{\kern0pt}designs\ {\isacharparenleft}{\kern0pt}\isakeyword{infixl}\ {\isachardoublequoteopen}{\isasymcong}\isactrlsub D{\isachardoublequoteclose}\ {\isadigit{5}}{\isadigit{0}}{\isacharparenright}{\kern0pt}\ \isakeyword{where}\ {\isachardoublequoteopen}{\isasymD}\ {\isasymcong}\isactrlsub D\ {\isasymD}{\isacharprime}{\kern0pt}\ {\isasymlongleftrightarrow}\isanewline {\isacharparenleft}{\kern0pt}{\isasymexists}\ {\isasympi}\ {\isachardot}{\kern0pt}\ design{\isacharunderscore}{\kern0pt}isomorphism\ {\isacharparenleft}{\kern0pt}points\ {\isasymD}{\isacharparenright}{\kern0pt}\ {\isacharparenleft}{\kern0pt}blocks\ {\isasymD}{\isacharparenright}{\kern0pt}\ {\isacharparenleft}{\kern0pt}points\ {\isasymD}{\isacharprime}{\kern0pt}{\isacharparenright}{\kern0pt}\ {\isacharparenleft}{\kern0pt}blocks\ {\isasymD}{\isacharprime}{\kern0pt}{\isacharparenright}{\kern0pt}\ {\isasympi}{\isacharparenright}{\kern0pt}{\isachardoublequoteclose}\medskip

\subsection{Graph Theory}

Graph theory proves an interesting case study when looking at extending the design hierarchy. As discussed in Sect.\ts2, simple graphs are designs. Can we link the design locale hierarchy to an existing formalisation, such as the general graph theory library in the AFP\@? This appears to present a number of challenges: (i) the graph theory library was developed in 2013 by a different author, (ii) the library includes digraphs, which are not designs, and (iii) the locale approach for graph theory uses records, which are not used for designs.

\begin{figure}[t]
    \centering
    \includegraphics[width=0.9\textwidth]{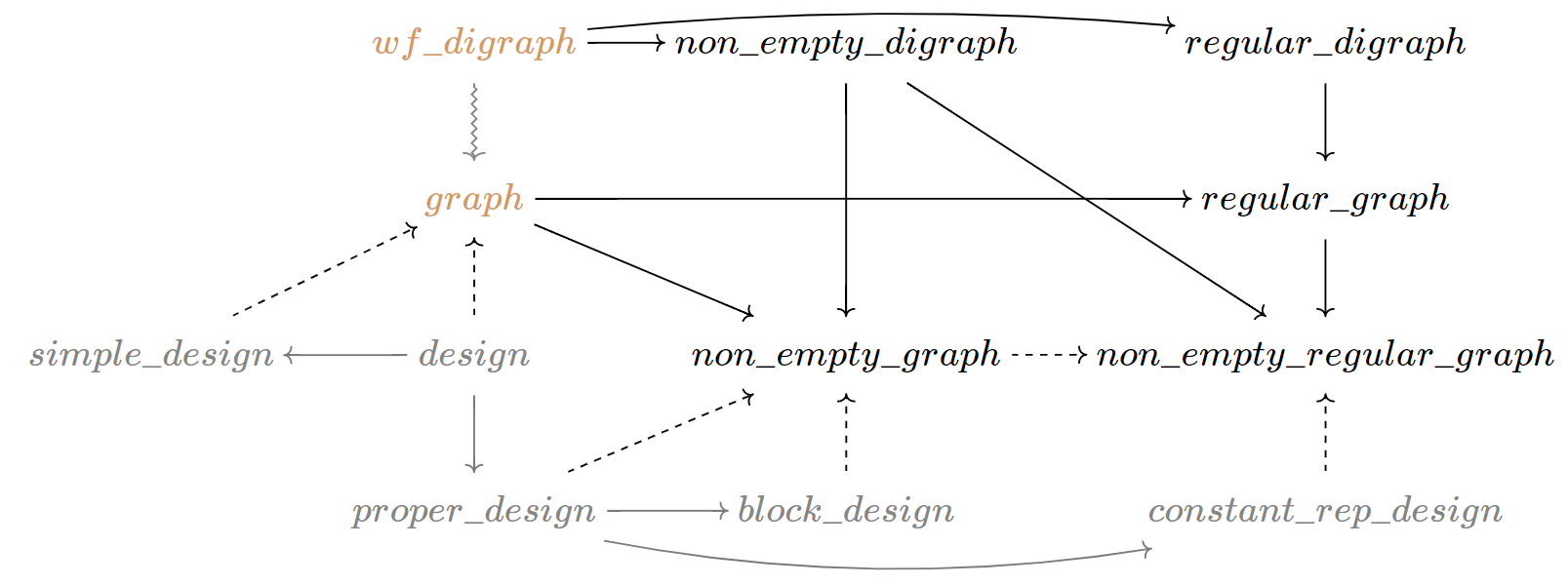}
    \caption{Interaction between Graph and Design Locales} \label{fig2}
\end{figure}

Despite these challenges, the flexibility of locales made it straightforward to prove that a simple graph is a design, as well as a number of other properties. Figure~\ref{fig2} shows the resulting links made between the design theory and graph theory locale developments, using sublocales.

To show that a graph is a design, we must convert the ordered edge representation to an unordered block. The \textit{arcs-blocks} definition manages the transformation within the \textit{graph} locale, which defines a simple graph by declaring the edge set to be symmetric without multiples or loops. A few lemmas ensure the translation is valid, from which it follows that a graph is a sublocale of a design.

\medskip\isacommand{sublocale}\isamarkupfalse%
\ graph\ {\isasymsubseteq}\ design\ {\isachardoublequoteopen}verts\ G{\isachardoublequoteclose}\ {\isachardoublequoteopen}arcs{\isacharunderscore}{\kern0pt}blocks{\isachardoublequoteclose}\medskip

Clearly, a non-empty graph is also a block design with $k = 2$, which is represented by another sublocale relationship. Additionally, we extended the existing graph theory library to define the concept of a \textit{regular-digraph} and \textit{regular graph}, which are of particular interest in design theory. In particular, a non-empty regular graph is a sublocale of a constant representation number design.

\medskip\isacommand{sublocale}\isamarkupfalse%
\ non{\isacharunderscore}{\kern0pt}empty{\isacharunderscore}{\kern0pt}reg{\isacharunderscore}{\kern0pt}graph{\isasymsubseteq}constant{\isacharunderscore}{\kern0pt}rep{\isacharunderscore}{\kern0pt}design\ {\isachardoublequoteopen}verts\ G{\isachardoublequoteclose}\ {\isachardoublequoteopen}arcs{\isacharunderscore}{\kern0pt}blocks{\isachardoublequoteclose}\ {\isasymr}

\section{The Modular Approach}

This paper has thus far demonstrated how locales can be used to build up an extensive hierarchy to formally reason on designs. This section discusses the benefits and limitations of the approach taken and recurring reasoning techniques. 

\subsection{The Formal Design Hierarchy}

\begin{figure}[b!]
    \centerline{\includegraphics[scale=0.7]{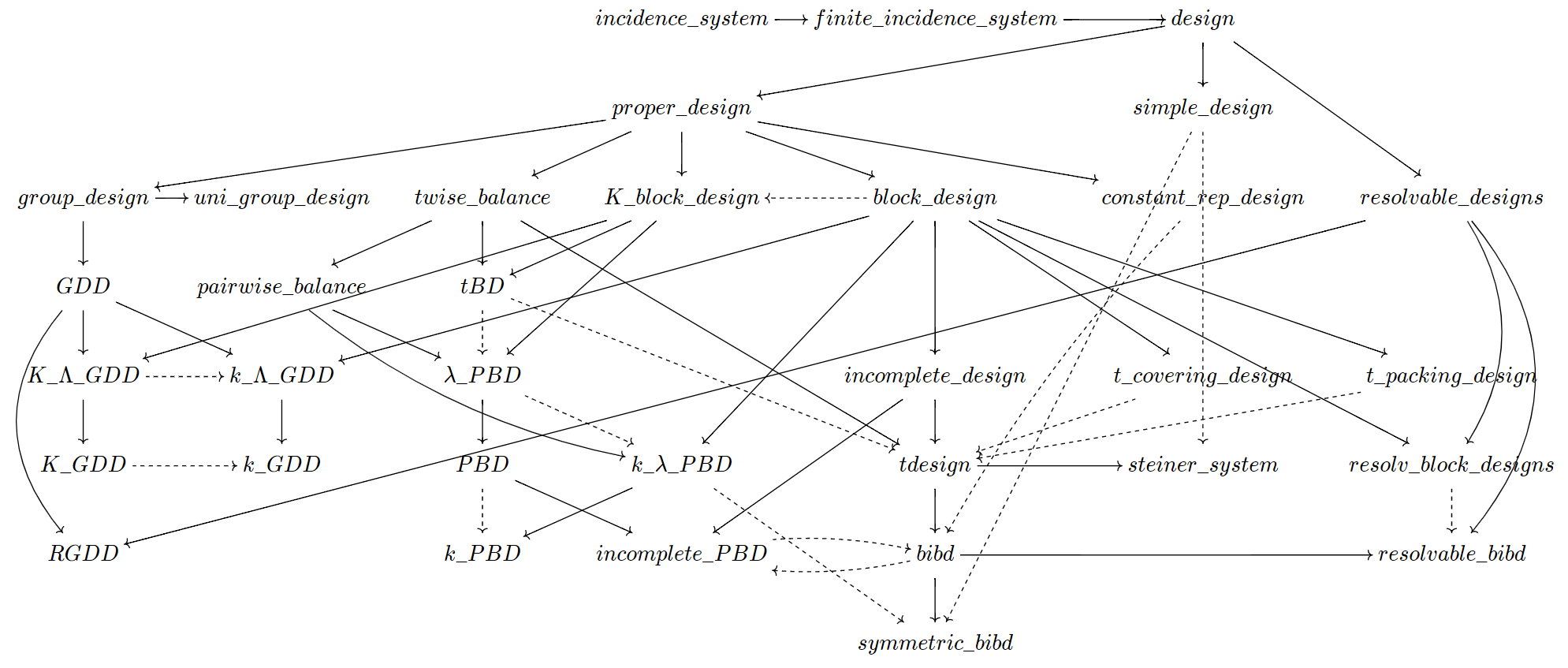}}
    \caption{Design Theory Locale Development} \label{fig3}
\end{figure}

This paper presents seemingly the first formalisation of design theory. As such, initial investigations focused on examining the approach taken by similar libraries on mathematical structures. There does exist a formalisation of Latin squares \cite{bentkampLatinSquare2015} in Isabelle. While these are a very specific type of design, their formalisation does not reflect this and is not extendable to designs generally. Rather, it highlights the need for flexibility when defining different design classes.

Type classes \cite{haftmannConstructiveTypeClasses2007} were briefly considered, however the constraints on parameters meant they didn't offer the same flexibility as locales. The ``record + locale'' approach first considered in (3.1) is based on Noschinski's graph theory library and the HOL-Algebra library. This approach uses a record to define structural elements and definitions, and locales for supporting concise syntax by parameter annotation \cite{ballarinExploringStructureAlgebra2020}. It was originally designed when definitions could not be declared within a locale and is still widely used. Changes to locales in 2009 \cite{haftmannLocalTheorySpecifications2009} however, enabled local theory specification, so definitions are now possible within a local context while still globally accessible. As such, structures can now be defined over a number of parameters within a locale without any noticeable limitations. This reduces the need for records and the required workarounds, while also simplifying notation and definitions for the structure.  

The small AFP development on matroids, another combinatorial structure, uses this more locale-centric approach \cite{keinholzMatroids2018}, but more interesting is Ballarin's take to formalising algebra~\cite{ballarinExploringStructureAlgebra2020}. He uses locales to define structures as well as operations and relationships on multiple instances of a locale, similar to the design isomorphism definition.

The final locale hierarchy of the design library can be seen in Fig.\ts\ref{fig3}, with some minor omissions. Figure~\ref{fig3} presents the numerous types of designs available in the formalisation and the complex inheritance network. The final formalisation defines 36 purely design related locales, as well as five new locales on graph theory. The larger graph theory library used only 21 locales.

\subsection{The Little Theories Approach}

Using the little theories approach, and drawing on ideas from the more radical tiny theories approach where suitable, each new locale declaration in the design library does some of three things: (i) combines multiple pre-existing locales, (ii) adds new parameters and assumptions related to a single new concept, or (iii) instantiates one or more parameters to a concrete value.  

This approach drew inspiration from both Noschinski and Ballarin~\cite{noschinskiGraphLibraryIsabelle2015,ballarinExploringStructureAlgebra2020}, and yielded a number of benefits, preventing unnecessary duplication when new designs were introduced. More importantly, it increased the flexibility and extensibility of the library. As can be seen from the case studies in Sect.\ts5 where the formalisation was extended, it was easy to integrate locales from the original hierarchy with new concepts. The sublocale command proved particularly useful in manipulating the hierarchy. Additionally, each extension took significantly less time than the original development due to the inherited material.

\subsection{Notational Benefits}

One of the key benefits that Ballarin discussed when comparing the locale-centric algebra approach with the existing library was notation, and its readability in comparison to a textbook \cite{ballarinExploringStructureAlgebra2020}. The locale-centric approach yields similar results for design theory. For example, in mathematical literature, a $t$-design is referred to as a $t$-$(v, k, \lambda_t)$-\textit{design}. In Isabelle, it would be represented by \textit{t-design $V$ $B$ $k$ $\Lambda_t$}, where $v$ can still be used to refer to the cardinality of $V$. 

In fact, all the usual single letter parameters are available with a design context, and definitions were done in locales where possible, thus the majority are simple and readable as is. The \textbf{for} command further increased readability by removing unnecessary parameters from specialisations. Overall, this results in concise notation both within a locale context and on instances of a locale, which should be readable for anyone familiar with design theory. Such notation also simplifies lemma statements, avoiding repeated assumptions, as well as proof goals. We expect that further extensions to different structures such as hypergraphs could benefit from locale notation features such as \textbf{rewrites}.

\subsection{Reasoning on Locales}

The flexibility of locales offers many benefits for reasoning. However, it is worth noting a number of proof patterns specific to working with locale definitions. 

Locales come with two proof tactics: \textit{unfold-locales}, which unfolds all the assumptions in the current context hierarchy, and \textit{intro-locales}, which unfolds to the axiomatic definitions of each locale in the current hierarchy.  The \textit{intro-locales} tactic was often used on proofs on the combine and multiple operations, which avoided the need to unfold all axioms for each proof.

Interpretations are likely the most powerful proof tool for locales, and can  decrease the complexity of proofs by providing an instance of a locale to refer to. The \textit{complement-bibd} lemma described in (4.5) is an example. 

\medskip\isacommand{lemma}\isamarkupfalse%
\ complement{\isacharunderscore}{\kern0pt}bibd{\isacharcolon}{\kern0pt}\ \isanewline
\ \ \isakeyword{assumes}\ {\isachardoublequoteopen}{\isasymk}\ {\isasymle}\ {\isasymv}\ {\isacharminus}{\kern0pt}\ {\isadigit{2}}{\isachardoublequoteclose}\ \isanewline
\ \ \isakeyword{shows}\ {\isachardoublequoteopen}bibd\ {\isasymV}\ {\isacharparenleft}{\kern0pt}complement{\isacharunderscore}{\kern0pt}blocks{\isacharparenright}{\kern0pt}\ {\isacharparenleft}{\kern0pt}{\isasymv}\ {\isacharminus}{\kern0pt}\ {\isasymk}{\isacharparenright}{\kern0pt}\ {\isacharparenleft}{\kern0pt}{\isasymb}\ {\isacharplus}{\kern0pt}\ {\isasymLambda}\ {\isacharminus}{\kern0pt}\ {\isadigit{2}}{\isacharasterisk}{\kern0pt}{\isasymr}{\isacharparenright}{\kern0pt}{\isachardoublequoteclose}\isanewline
\isacommand{proof}\isamarkupfalse%
\ {\isacharminus}{\kern0pt}\isanewline
\ \ \isacommand{interpret}\isamarkupfalse%
\ des{\isacharcolon}{\kern0pt}\ incomplete{\isacharunderscore}{\kern0pt}design\ {\isasymV}\ {\isachardoublequoteopen}{\isacharparenleft}{\kern0pt}complement{\isacharunderscore}{\kern0pt}blocks{\isacharparenright}{\kern0pt}{\isachardoublequoteclose}\ {\isachardoublequoteopen}{\isacharparenleft}{\kern0pt}{\isasymv}\ {\isacharminus}{\kern0pt}\ {\isasymk}{\isacharparenright}{\kern0pt}{\isachardoublequoteclose}\isanewline \ \ \ \isacommand{using}\isamarkupfalse%
\ assms\ complement{\isacharunderscore}{\kern0pt}incomplete\ \isacommand{by}\isamarkupfalse%
\ blast\isanewline
\ \ \isacommand{show}\isamarkupfalse%
\ {\isacharquery}{\kern0pt}thesis\ \isacommand{proof}\isamarkupfalse%
\ {\isacharparenleft}{\kern0pt}unfold{\isacharunderscore}{\kern0pt}locales{\isacharcomma}{\kern0pt}\ simp{\isacharunderscore}{\kern0pt}all{\isacharparenright}{\kern0pt}\isanewline
\ \ \ \ \isacommand{show}\isamarkupfalse%
\ {\isachardoublequoteopen}{\isadigit{2}}\ {\isasymle}\ des{\isachardot}{\kern0pt}{\isasymv}{\isachardoublequoteclose}\ \isacommand{using}\isamarkupfalse%
\ assms\ block{\isacharunderscore}{\kern0pt}size{\isacharunderscore}{\kern0pt}t\ \isacommand{by}\isamarkupfalse%
\ linarith\ \isanewline
\ \ \ \ \isacommand{show}\isamarkupfalse%
\ {\isachardoublequoteopen}{\isasymAnd}ps{\isachardot}{\kern0pt}\ ps\ {\isasymsubseteq}\ {\isasymV}\ {\isasymLongrightarrow}\ card\ ps\ {\isacharequal}{\kern0pt}\ {\isadigit{2}}\ {\isasymLongrightarrow}\isanewline 
\ \ \ \ points{\isacharunderscore}{\kern0pt}index\ {\isacharparenleft}{\kern0pt}complement{\isacharunderscore}{\kern0pt}blocks{\isacharparenright}{\kern0pt}\ ps\ {\isacharequal}{\kern0pt}\ {\isasymb}\ {\isacharplus}{\kern0pt}\ {\isasymLambda}\ {\isacharminus}{\kern0pt}\isanewline
\ \ \ \ {\isadigit{2}}\ {\isacharasterisk}{\kern0pt}\ {\isacharparenleft}{\kern0pt}{\isasymLambda}\ {\isacharasterisk}{\kern0pt}\ {\isacharparenleft}{\kern0pt}des{\isachardot}{\kern0pt}{\isasymv}\ {\isacharminus}{\kern0pt}\ {\isadigit{1}}{\isacharparenright}{\kern0pt}\ div\ {\isacharparenleft}{\kern0pt}{\isasymk}\ {\isacharminus}{\kern0pt}\ {\isadigit{1}}{\isacharparenright}{\kern0pt}{\isacharparenright}{\kern0pt}{\isachardoublequoteclose}\
\ \isacommand{using}\isamarkupfalse%
\ complement{\isacharunderscore}{\kern0pt}bibd{\isacharunderscore}{\kern0pt}index\ \isacommand{by}\isamarkupfalse%
\ simp\isanewline
\ \ \ \ \isacommand{show}\isamarkupfalse%
\ {\isachardoublequoteopen}{\isadigit{2}}\ {\isasymle}\ des{\isachardot}{\kern0pt}{\isasymv}\ {\isacharminus}{\kern0pt}\ {\isasymk}{\isachardoublequoteclose}\ \isacommand{using}\isamarkupfalse%
\ assms\ block{\isacharunderscore}{\kern0pt}size{\isacharunderscore}{\kern0pt}t\ \isacommand{by}\isamarkupfalse%
\ linarith\ \isanewline
\ \ \isacommand{qed}\isamarkupfalse%
\medskip

The \textbf{interpret} command yields an instance of an incomplete_design with the complement parameters. To prove the conclusion, after applying \textit{unfold_locales} and simplification, we get three sub-goals instead of the 10 \textit{unfold_locales} gives without interpretation. This is both easy to approach and read.

Another useful pattern that assists automation is defining custom introduction rules, particularly around reverse sublocale relationships. For example, an introduction rule can be proven stating that parameters which satisfy the axioms of $t$-covering and $t$-packing designs also satisfy the $t$-design axioms. Ballarin's functor pattern \cite{ballarinExploringStructureAlgebra2020}, which connects two linear locale hierarchies related by a functor using a series of sublocale declarations, is also used in the formalisation. An example of this can be seen from the GDD variations in Fig.\ts\ref{fig3}.

Lastly, we also note the ease of reasoning on multiple labelled instances of a locale, within another locale. The prime example of this is in the \textit{design-isomorphism} theory. This is a technique that could be explored further for other operations and relationships, such as the concept of \textit{sub-designs}.

\subsection{Limitations}

A few limitations of the locale-centric approach to mathematics are worth noting. First, locale specifications were not designed to be used extensively outside of the locale. However, the approach requires this, which particularly causes issues with sublocales. A sublocale proof does not generate any additional facts, and as such cannot be referenced; to reference this relationship for reasoning outside of the locale, one must define a separate lemma with a nearly identical proof.

While the \textbf{interpret} command within proofs is incredibly useful, it would be beneficial to see extensions to locale proof tactics to aid automation and proof structure. Many \textbf{interpret} declarations are trivial consequences of known facts, but they must be written out in full.

Lastly, the little theories approach can cause locale hierarchies to become complex. We need ways to keep track of relationships between locales during development. In particular, sublocale relationships must be maintained and added carefully when frequently combining locales at different levels in the hierarchy.

\section{Conclusion and Future Work}

Through the use of locales, this paper demonstrates how the complex hierarchy of design theoretic structures can be formalised in a proof assistant, presenting the first such formalisation for this field. It is intended that this library will be used to further explore some of the unique challenges combinatorial proofs currently pose to formalisation. The locale-centric modular approach discussed has proven to be an effective method of concisely and accurately defining numerous fundamental properties and classes of designs, and reasoning on key theorems and inheritance relationships. Additionally, the case studies presented in Sect.\ts5 demonstrates the formalisation's flexibility and extensibility for future work on design theory and other related combinatorial structures, fulfilling the aim of establishing a general adaptable library for designs. This library will be made available in full through the Isabelle Archive of Formal Proofs. Beyond the obvious potential to continue formalising new classes of designs, other future work includes further exploring locale-centric proof techniques and improvements, experimenting with links to hypergraphs, and investigating the formalisation of theorems on designs which involve more advanced and varied proof techniques.

%
%
%
\bibliographystyle{splncs04}
\bibliography{cicm2021}
\end{document}